\documentclass[preprint,
 amsmath,amssymb]{revtex4-1}
\usepackage{graphicx}% Include figure files
\usepackage{dcolumn}% Align table columns on decimal point
\usepackage{bm}%
\usepackage{epstopdf}
\usepackage[normalem]{ulem}
\usepackage[usenames]{color}

\begin{document}

\title{Low-field microwave absorption and magnetoresistance in iron nanostructures grown by electrodeposition on n-type lightly-doped silicon substrates
}

\author{J. F. Felix}
\affiliation{ Universidade Federal de Vi\c cosa-UFV, Departamento de F\'{i}sica, 36570-900, Vi\c cosa, MG, Brazil }
\affiliation{Universidade de Bras\'{i}lia-UnB, Instituto de F\'{i}sica, N\'{u}cleo de  F\'{i}sica Aplicada, 70910-900, Bras\'{i}lia, DF, Brazil }

\author{L. C. Figueiredo}
\affiliation{Universidade de Bras\'{i}lia-UnB, Instituto de F\'{i}sica, N\'{u}cleo de  F\'{i}sica Aplicada, 70910-900, Bras\'{i}lia, DF, Brazil }

\author{J. B. S. Mendes} 
\affiliation{  Universidade Federal de Vi\c cosa-UFV, Departamento de F\'{i}sica, 36570-900, Vi\c cosa, MG, Brazil }

\author{P. C. Morais}
\affiliation{Universidade de Bras\'{i}lia-UnB, Instituto de F\'{i}sica, N\'{u}cleo de  F\'{i}sica Aplicada, 70910-900, Bras\'{i}lia, DF, Brazil }
\affiliation{Huazhong University of Science and Technology, School of Automation, 430074, Wuhan, China}

\author{C. I. L. de Araujo}
\email{dearaujo@ufv.br}
\affiliation{ Universidade Federal de Vi\c cosa-UFV, Departamento de F\'{i}sica, 36570-900, Vi\c cosa, MG, Brazil }

\date{\today}

\begin{abstract}
\subsection{Objective}
In this study we investigate magnetic properties, surface morphology and crystal structure in iron nanoclusters electrodeposited on lightly-doped (100) n-type silicon substrates. Our goal is to investigate the spin injection and detection in the Fe/Si lateral structures.
\subsection{Methods}
The samples obtained under electric percolation were characterized by magnetoresistive and magnetic resonance measurements with cycling the sweeping applied field in order to understand the spin dynamics in the as-produced samples. 
\subsection{Results}
The observed hysteresis in the magnetic resonance spectra, plus the presence of a broad peak in the non-saturated regime confirming the low field microwave absorption (LFMA), were correlated to the peaks and slopes found in the magnetoresistance curves. 
\subsection{Conclusion}
The results suggest long range spin injection and detection in low resistive silicon and the magnetic resonance technique is herein introduced as a promising tool for analysis of electric contactless magnetoresistive samples.

\end{abstract}

\pacs{Valid PACS appear here}% PACS, the Physics and Astronomy
\maketitle

\section{Introduction}
The intrinsic spin density imbalance between minority and majority carriers in ferromagnetic materials\cite{fert} allows for its application as spin polarized electrodes in spintronic devices with main applications in magnetic
sensors, as for instance spin valves\cite{dieny} and magnetic tunnel junctions\cite{moodera1,moodera2} or in fundamental 
investigations of spin logic devices\cite{data, appelbaum, dearaujo}. Among the ferromagnetic materials iron stands out due to its high saturation magnetization, high curie temperature\cite{Komogortsev}, enhanced spin asymmetry\cite{Meservey} and low lattice mismatch with spintronic suitables substrates, as for instance the case of (100) silicon
and magnesium oxide\cite{review}. Another iron-related feature is the possibility of its oxidation to the half metallic magnetite\cite{zolda}, to which is assigned full bulk spin polarization and low impedance mismatch with semiconductor interfaces, both fundamental characteristics for good spin polarized current injection\cite{molenkamp}.\\
Structural and morphological characteristics of ferromagnetic materials can affect their magnetic response, $e.g.$ typical magnetic anisotropy\cite{phuoc} or inhomogeneous magnetostatic field near a ferromagnetic interface\cite{dash}. In order to investigate such effects static and dynamic magnetization characterizations are often carried out. While static magnetization data provides physical properties of magnetic nanostructures such as saturation magnetization, remanence and coercivity\cite{Wu}, dynamic magnetization measurements can be used to assess magnetic anisotropy, spin splitting g-factor, high-frequency loss magnetization relaxation, and magnetic inhomogeneity\cite{Mendes,Fermin,Barsukov,Rodriguez-Suarez}. Dynamic magnetization measurements can be performed via magnetic resonance experiments, which are characterized by the onset of absorption peaks when the incident microwave frequency matches the spin precession frequency.\\
In nanostructured and ferrite-based materials (e.g. nanoparticles and microwires)\cite{np, nw, ferrites} a low-field microwave absorption (LFMA) band is usually observed in non-saturated remnant states, which is mainly attributed to the high random anisotropy found in these systems\cite{Kim}. In the present study we report on the use of the electrodeposition technique to fabricate iron-based nanoclusters of ferromagnetic electrodes on lightly-doped n-type silicon substrates. This technique is one of the most suitable to fabricate ferromagnetic deposits on silicon substrates, particularly due to the low contamination profile achieved, large scale production availability, and low cost. Our goal is to carry on the electrodeposition of iron on high resistive substrates, characterize the magnetic properties of the as-deposited materials and investigate the spin injection and transport properties in the silicon using local magnetoresistive measurements in the non-electrically percolated iron nanoclusters regime.

\section{Materials and methods}

The electrodeposition of the ferromagnetic nanoclusters was carried out on lightly-doped 1k $\Omega$.cm n-type (100) silicon substrate in a delimited area of about 0.5 ${\rm cm^{2}}$. Preparation of the bath solution for the electrodeposition, (containing 0.5M Fe${\rm}_{2}$SO${\rm}_{4}$ and 0.5M NaSO${\rm}_{4}$), followed the protocol described in the literature \cite{zangari} whereas the pH = 2.5 of the bath solution was controlled by sulfuric acid addition. 
A set of samples, with different deposition time, were produced in the potentiostatic mode using a short nucleation time process (30 sec) under low potential (of -1.2 V  \textit{vs} calomel reference electrode), following by 15 to 60 sec of deposition under -1.8 V.
The electrodeposition time and the corresponding deposit resistance were monitored in order to find out the electrical percolation condition. Then, the investigation was focused on the samples obtained under this particular electrodeposition time, in order to allow for the current flowing through the subsequent Fe/Si/Fe lateral structures.\\
The crystallographic structure of the deposits was assessed by X-ray diffraction (XRD) measurements. XRD patterns were recorded using the Bruker D8 Discover Diffractometer equipped with the Cu K$\alpha$ radiation ($\lambda$ = 1.5418 $\AA$). The diffraction patterns were obtained at angles between 35${\rm ^{o}}$ and 90${\rm ^{o}}$ (2$\theta$). The films morphology were investigated by field-emission scanning electron microscopy (FEG-SEM), whereas the magnetization loops and the magnetic resonance spectra were recorded using a vibrating sample magnetometer and a X-band spectrometer (Bruker EMX PremiumX equipped with the ER 4102ST resonator), respectively.
For the magnetoresistive measurements (2x4 mm $GaIn$ alloy) two probe ohmic contacts (separated by 2 mm) were painted on the samples and connected by conventional Copper wires.

\section{Results and discussions}

\textbf{Figures 1a} and \textbf{1b} present the images of the ferromagnetic samples obtained after shorter (15 sec) and longer (60 sec) deposition times, respectively. 
%-------------------------------------------------------------------------------------------
\begin{figure}[!h]
		\center
		\includegraphics[width=8.5cm]{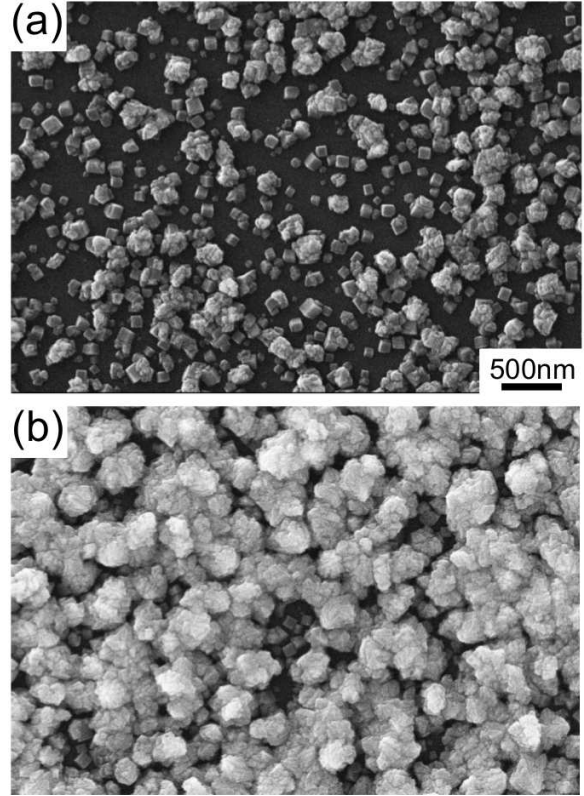}
		\caption {Scanning electron microscopy micrograph of the iron samples obtained with electrodeposition time of a) 15 sec and b) 60 sec (scale bar = 500 nm). The images reveal isolated clusters at the shortest deposition time (15 sec) and its evolution to a percolated film at the longest deposition time (60 sec).}
		\label{fig:fig1}
	\end{figure}	
%-------------------------------------------------------------------------------------------
The SEM image presented in the \textbf{Figure 1a} shows the shortest deposition time (15 sec) sample, revealing non-percolated cubic-shaped nanoclusters with good homogeneity in size, more likely achieved due to the initial nucleation process performed at low potential (-1.2 V). \textbf{Figure 1b}, shows the SEM image of the sample obtained after the longest deposition time (60 sec), revealing a percolated thin film evolved from the cubic-shaped morphology towards a columnar faceted structure with grain size of about 500 nm.\\ 
Due to the low iron-content in the as-deposited samples improved signal detection and/or peak-to-noise ratio associated to the iron reflections was achieved using low incident angle XRD. Therefore, as presented in  \textbf{Figure 2}, the XRD spectrum was recorded using the iron sample deposited during 60 sec where the three XRD peaks associated with the (011), (002) and (112) crystal planes can be observed. The peaks observed in the XRD pattern correspond to the bbc iron structure which is indexed on the basis ICSD data card  180969. Moreover, the 60 sec as-deposited iron sample exhibited preferential orientation along the (011) crystal plane.\\
%-------------------------------------------------------------------------------------------
\begin{figure}
		\center
		\includegraphics[width=8.5cm]{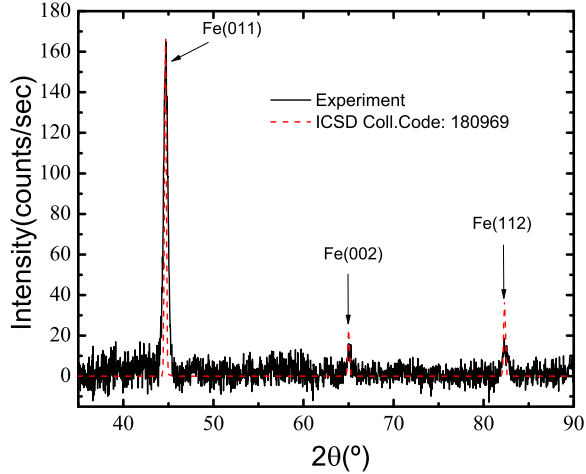}
		\caption{XRD spectrum showing the out-of-plane orientation of the iron sample grown on the Si (100) substrate. In order to suppress the strong substrate XRD peaks the spectrum was recorded with 1${\rm ^{o}}$ offset.}
		\label{fig:fig1}
	\end{figure}	
%-------------------------------------------------------------------------------------------  
\textbf{Figure 3a} shows the magnetic resonance spectra recorded from the as-deposited iron sample with the deposition time set to 15 sec. The magnetic resonance spectra were recorded in the parallel (in-plane) as well as in the perpendicular (out-of-plane) orientation of the applied magnetic field with respect to the substrate surface. In order to assess the samples magnetic hysteresis the magnetic resonance spectra were recorded while cycling the applied magnetic field. Notice the presence of the LFMA peak in the magnetic resonance spectra recorded in the parallel orientation (0${\rm ^{o}}$), with a clear hysteretic behavior while cycling the applied magnetic field. additionally, in the parallel orientation a single broad absorption line is observed over the whole range of sweeping field. Actually, the observed broadening can be attributed to the in-plane (0${\rm ^{o}}$) sample polycrystallinity and shape anisotropy dominating the as-deposited iron nanoclusters. Notice from the measurements the maximum coercivity is observed for the perpendicular orientation of applied magnetic field (90${\rm ^{o}}$), meaning the orientation of the films hard-axis. \textbf{Figure 3b} shows the quenching of the broad magnetic resonance component while performing the angular variation measurements from 0${\rm ^{o}}$ (in-plane) to 90${\rm ^{o}}$ (out-of-plane). \\
%-------------------------------------------------------------------------------------------
\begin{figure}
		\includegraphics[width=8.5cm]{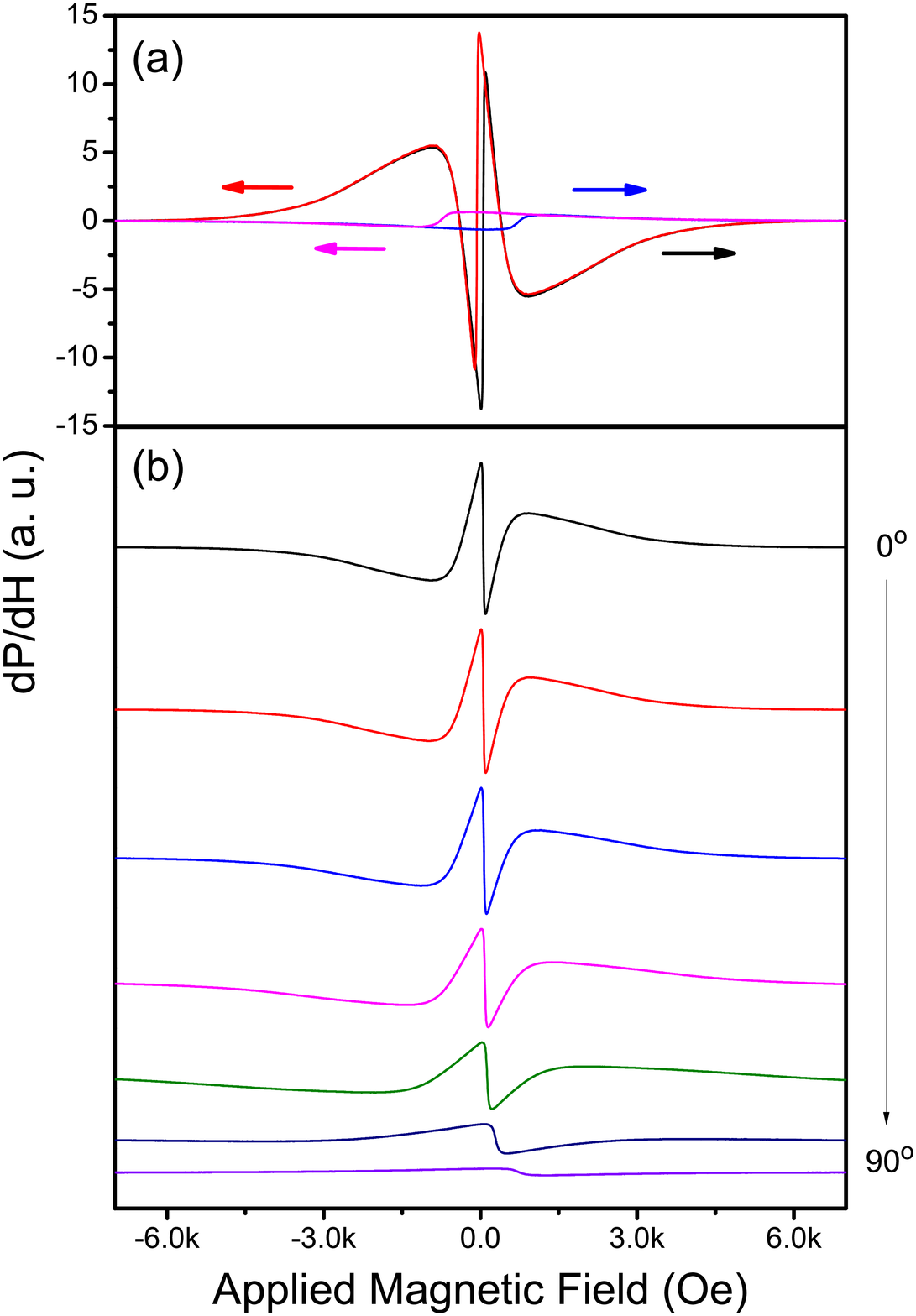}
		\caption{Room temperature magnetic resonance spectra recorded using the 30 sec deposition time sample while a) cycling the applied magnetic field in the parallel (red and black solid line) and perpendicular (blue and pink solid line) orientation, showing the hysteretic behavior of the LFMA line and b) performing the angular variation from 0${\rm ^{o}}$ towards 90${\rm ^{o}}$ orientation, revealing the quenching of the broad magnetic resonance feature.}
		\label{fig:fig1}
	\end{figure}	
%-------------------------------------------------------------------------------------------
In order to investigate the samples behavior in regard to the spin injection, transport and detection we have performed local magnetoresistive measurements. 
\textbf{Figures 4a} and \textbf{4b} show the magnetoresistance measurements recorded from the as-deposited iron sample with the deposition time set to 15 sec with the external magnetic field applied in the in-plane configuration but longitudinally and transversely oriented with respect to the applied DC electric current (1 mA), respectively. \textbf{Figure 4c} presents the magnetoresistance measurement recorded from the same sample (15 sec deposition time) with the external magnetic field applied in the out-of-plane configuration and therefore perpendicular to the applied DC electric current. The isotropic magnetoresistance response of the investigated sample, similar to the behavior observed in the ferromagnetic/non-magnetic multilayers \cite{multilayer} and granular magnetic samples\cite{clusters1,clusters2,clusters3}, suggests successful spin injection and detection once at low external magnetic field the highest resistance can be associated with the random in-plane clusters magnetization. The random in-plane cluster magnetization increases the resistance of both majority and minority spin channels\cite{mott} whereas the cluster ferromagnetic orientation by the external applied magnetic field decreases the measured resistance and enables electric spin polarized current flowing from iron clusters throughout the silicon substrate. In comparison with the typical values found for spin valves or tunnel junctions the low value of the herein measured magnetoresistance ($\sim$0.1\%), calculated from the difference between the maximum and minimum measured resistances, is attributed to the large spatial separation among clusters which is about the same order of magnitude of the values predicted for the silicon spin diffusion lengths\cite{dearaujo, ref1,ref2}, high impedance mismatch in the ferromagnetic/semiconductor interface due to the high resistivity of the silicon substrate and higher rates of spin mixing in the thick depletion layer present in highly-resistive semiconductors\cite{jansen}.\\
%------------------------------------------------------------------------------------------
\begin{figure}
		\center
		\includegraphics[width=8.5cm]{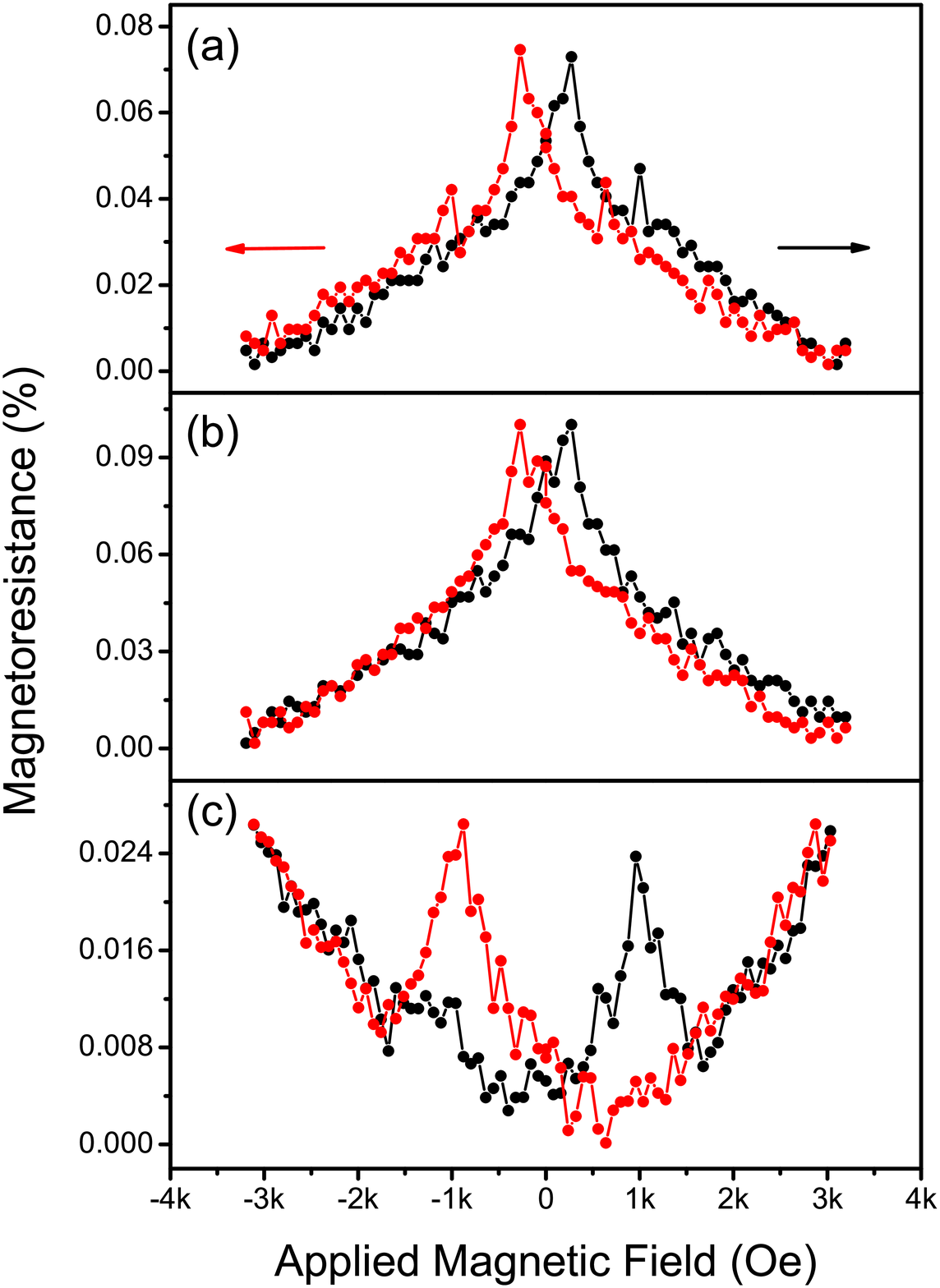}
		\caption{Room temperature magnetoresistance measurements recorded using the 15 sec deposition time sample with the (a) in-plane external magnetic field applied longitudinally to the electric current, (b) in-plane external magnetic field applied transversally to the electric current, and (c) external magnetic field applied perpendicularly to the electric current (out-of-plane).}
		\label{fig:fig1}
	\end{figure}	
%-------------------------------------------------------------------------------------------
For comparison, (\textbf{Figure 5}) presents the magnetic hysteresis cycle (\textbf{Figure 5a}), the cycling magnetic resonance spectra (\textbf{Figure 5b}) and the magnetoresistance (\textbf{Figure 5c}) data recorded from the as-deposited iron sample (30 sec deposition time) with the external magnetic field applied in the in-plane configuration and transversely oriented with respect to the applied $DC$ electric current (1 mA). The blue vertical dotted lines (see Figure 5) clearly show that different techniques provide consistent values for the sample coercivity. In addition, comparison of the data presented in \textbf{Figure 5} shows that the broad magnetic resonance line absorption vanishes at the field range where saturation in both magnetoresistance and magnetization occurs, thus supporting the model picture for the origin of the broad magnetic resonance absorption arising from the nonsaturated anisotropic magnetization. The magnetoresistance measurements suggest the increase in the current spin polarization while under the external applied magnetic field, random magnetization with high rate of spin mixing and, consequently, low spin polarization at zero applied field. Ferromagnetic switch around the coercive field matches the observed magnetoresistance peaks. The origin of the magnetoresistance effect mentioned above is closely related to the field dependence of the spin dynamics, as observed in the magnetic resonance measurements\cite{paper1, paper2}. This finding allows contactless analysis of the magnetoresistance effect using magnetic resonance measurements, preventing realization of electric contacts for magnetoresistive measurements in all investigated samples.
%-----------------------------------------------------------------------------------------
\begin{figure}
		\center
		\includegraphics[width=8.0cm]{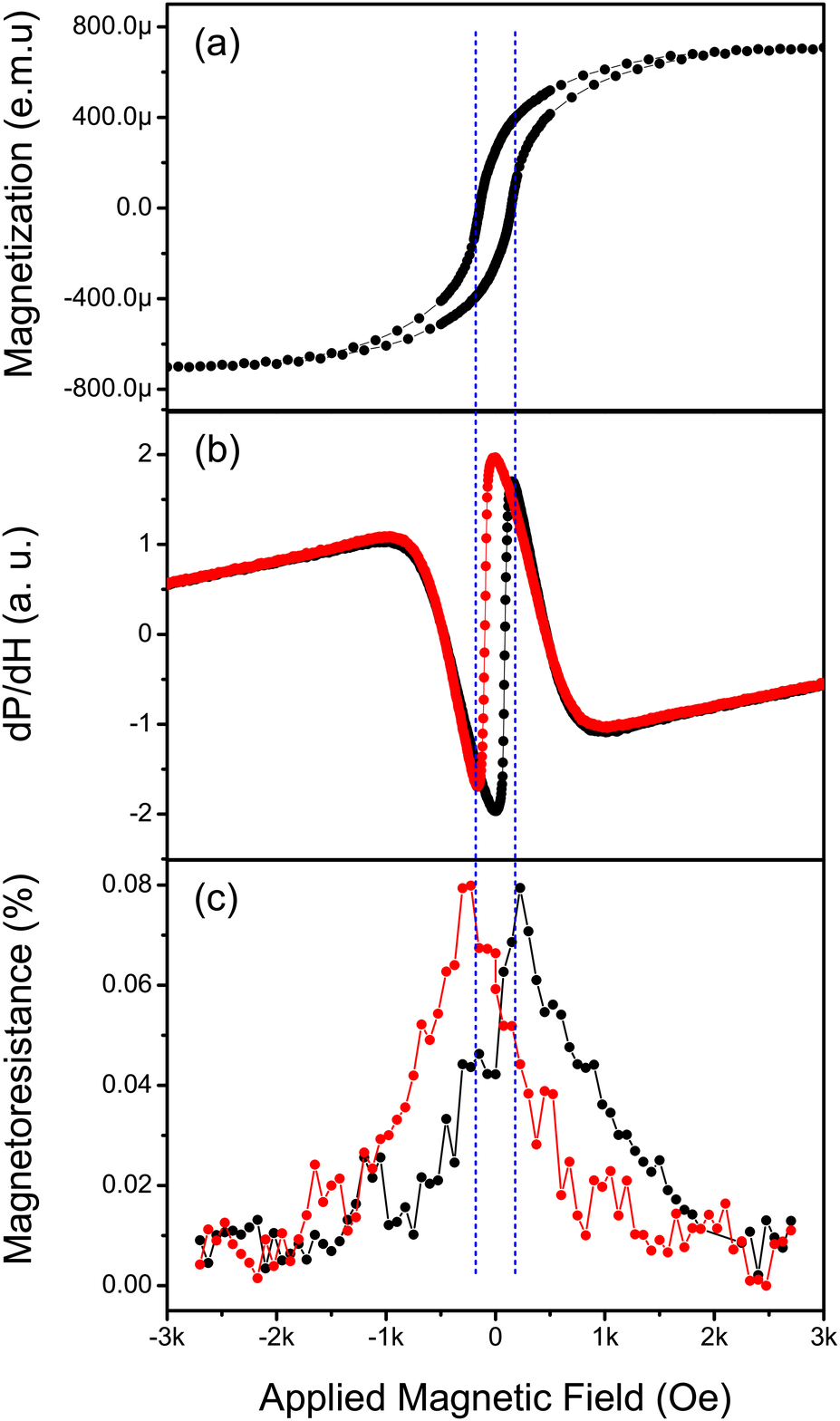}
		\caption{Comparison of the features among room temperature a) magnetic hysteresis cycle, b) magnetic resonance absorption and c) magnetoresistance, measured in the 30 sec deposition time sample. The external magnetic field was applied in the in-plane orientation with respect to the silicon substrate.}
		\label{fig:fig1}
	\end{figure}	
%------------------------------------------------------------------------------------------
\section{Conclusion}
In conclusion, the electrodeposition technique was successfully used to grow iron-based ferromagnetic ordered nanostructures on high resistive silicon substrates. Based on the magnetic properties herein reported we envisage that the iron/silicon interface in the as-growth nanostructures might have promising application in future spintronics logic devices. Particularly interesting is the low impurity content achieved while using the electrodeposition technique and lightly-doped silicon, leading to low spin mixing centers and promising spin polarization improvement with possibility of half metallic magnetite formation via oxidation. Additionally, the magnetoresistivity characterization assessed via magnetic resonance is herein introduced as a very promising non-contact technique for easy investigation in this material system.\\

\section{acknowledgments}
The authors would like to thank the Brazilian funding agencies through grants from CAPES, FAPEMIG and CNPq for supporting this project, particularly under the research grant number 472492/2013-6.\\

\end{document}